\documentclass[12pt]{article}
\usepackage{amssymb,amsfonts}
\newlength{\extraspace}
\newlength{\extraspaces}
\newcommand{\be}{\begin{equation}
\addtolength{\abovedisplayskip}{\extraspaces}
\addtolength{\belowdisplayskip}{\extraspaces}
\addtolength{\abovedisplayshortskip}{\extraspace}
\addtolength{\belowdisplayshortskip}{\extraspace}}
\newcommand{\ee}{\end{equation}}

\newcommand{\ba}{\begin{eqnarray}
\addtolength{\abovedisplayskip}{\extraspaces}
\addtolength{\belowdisplayskip}{\extraspaces}
\addtolength{\abovedisplayshortskip}{\extraspace}
\addtolength{\belowdisplayshortskip}{\extraspace}}
\newcommand{\ea}{\end{eqnarray}}

\newcommand{\newsection}[1]{
\vspace{15mm}
\pagebreak[3]
\addtocounter{section}{1}
\setcounter{equation}{0}
\setcounter{subsection}{0}
\setcounter{footnote}{0}
\begin{flushleft}
{\large\bf \thesection. #1}
\end{flushleft}
\nopagebreak
\medskip
\nopagebreak}

\newcommand{\Tr}{{\rm Tr}}
\newcommand{\La}{\mathcal{L}}

\newcommand{\qu} {\mathbf{Q}}
\newcommand{\I}{\textbf{I}}
\newcommand{\Ucr}{U^{\dagger}}
\newcommand{\Xcr}{X^{\dagger}}

\newcommand{\grup}{U(3)\otimes U(3)}
\newcommand{\sgru}{SU(L)\otimes SU(L)}

\newcommand{\csti}{F^{2}_{\pi}+3F^{2}_{X}}

\newcommand{\eps}{\varepsilon^{\mu\nu\rho\sigma}}

\begin{document}

\addtolength{\baselineskip}{.8mm}

{\thispagestyle{empty}
\noindent \hspace{1cm}  \hfill IFUP--TH/2003--42 \hspace{1cm}\\
\mbox{}                 \hfill October 2003 \hspace{1cm}\\

\begin{center}
\vspace*{1.0cm}
{\large\bf The effects of an extra $U(1)$ axial condensate on}\\
{\large\bf the radiative decay $\eta' \to \gamma\gamma$ at finite
temperature} \\
\vspace*{1.0cm}
{\large E. Meggiolaro}\\
\vspace*{0.5cm}{\normalsize
{Dipartimento di Fisica, \\
Universit\`a di Pisa, \\
Via Buonarroti 2, \\
I--56127 Pisa, Italy.}}\\
\vspace*{2cm}{\large \bf Abstract}
\end{center}

\noindent
Supported by recent lattice results, we consider a scenario in which a
$U(1)$--breaking condensate survives across the chiral transition in QCD.
This scenario has important consequences on the pseudoscalar--meson sector,
which can be studied using an effective Lagrangian model.
In particular, generalizing the results obtained in a previous paper
(where the zero--temperature case was considered), we study the effects of
this $U(1)$ chiral condensate on the radiative decay $\eta' \to \gamma\gamma$
at finite temperature.

\vspace{0.5cm}
\noindent
(PACS codes: 12.38.Aw, 12.39.Fe, 11.15.Pg, 11.30.Rd)
}
\vfill\eject

\newsection{Introduction}

\noindent
There are evidences from some lattice results
\cite{Bernard-et-al.97,Karsch00,Vranas00} that a new $U(1)$--breaking
condensate survives across the chiral transition at $T_{ch}$, staying
different from zero up to a temperature $T_{U(1)} > T_{ch}$.
$T_{U(1)}$ is, therefore, the temperature at which the $U(1)$ axial symmetry
is (effectively) restored, meaning that, for $T>T_{U(1)}$, there are no
$U(1)$--breaking condensates.
This scenario has important consequences on the pseudoscalar--meson sector,
which can be studied using an effective Lagrangian model
\cite{EM1994a,EM1994b,EM1994c,EM2002}, including also the new $U(1)$ chiral
condensate. This one has the form $C_{U(1)} = \langle {\cal O}_{U(1)} \rangle$,
where, for a theory with $L$ light quark flavours, ${\cal O}_{U(1)}$ is a
$2L$--fermion local operator that has the chiral transformation properties of
\cite{tHooft76}:\footnote{Throughout this paper we use the following notations
for the left--handed and right--handed quark fields:
$q_{L,R} \equiv {1 \over 2} (1 \pm \gamma_5) q$,
with $\gamma_5 \equiv -i\gamma^0\gamma^1\gamma^2\gamma^3$.}
\be
{\cal O}_{U(1)} \sim \displaystyle{{\det_{st}}(\bar{q}_{sR}q_{tL})
+ {\det_{st}}(\bar{q}_{sL}q_{tR}) },
\label{Ou1}
\ee
where $s,t = 1, \ldots ,L$ are flavour indices; the colour indices [not
explicitly indicated in Eq. (\ref{Ou1})] are arranged in such a way that:
{\it i)} ${\cal O}_{U(1)}$ is a colour singlet, and {\it ii)}
$C_{U(1)} = \langle {\cal O}_{U(1)} \rangle$ is a {\it genuine} $2L$--fermion
condensate, i.e., it has no {\it disconnected} part proportional to some
power of the quark--antiquark chiral condensate $\langle \bar{q} q \rangle$
(see Refs. \cite{EM1994c,EM2002,EM1995}).

The low--energy dynamics of the pseudoscalar mesons, including the effects due
to the anomaly, the $q\bar{q}$ chiral condensate and the new $U(1)$ chiral
condensate, can be described, in the limit of large number $N_c$ of colours,
and expanding to the first order in the light quark masses, by an effective
Lagrangian written in terms of the topological charge density $Q$, the mesonic
field $U_{ij} \sim \bar{q}_{jR} q_{iL}$ (up to a multiplicative constant) and
the new field variable $X \sim {\det} \left( \bar{q}_{sR} q_{tL} \right)$
(up to a multiplicative constant), associated with the new $U(1)$ condensate
\cite{EM1994a,EM1994b,EM1994c,EM2002}:
\ba
\lefteqn{
{\cal L}(U,U^\dagger ,X,X^\dagger ,Q)
= {1 \over 2}\Tr(\partial_\mu U\partial^\mu U^\dagger )
+ {1 \over 2}\partial_\mu X\partial^\mu X^\dagger } \nonumber \\
& & -V(U,U^\dagger ,X,X^\dagger)
+ {i \over 2}\omega_1 Q \Tr(\ln U - \ln U^\dagger) \nonumber \\
& & + {i \over 2}(1-\omega_1)Q(\ln X-\ln X^\dagger) + {1 \over 2A}Q^2,
\label{lagrangian}
\ea
where the potential term $V(U,U^{\dagger},X,X^{\dagger})$ has the form:
\ba
\lefteqn{
V(U,U^\dagger ,X,X^\dagger )
= {\lambda_{\pi}^2 \over 4} \Tr[(U^\dagger U
-\rho_\pi {\bf I})^2] +
{\lambda_X^2 \over 4} (X^\dagger X-\rho_X )^2 } \nonumber \\
& & -{B_m \over 2\sqrt{2}}\Tr(MU+M^\dagger U^\dagger)
-{c_1 \over 2\sqrt{2}}[\det(U)X^\dagger + \det(U^\dagger )X].
\label{potential}
\ea
$M={\rm diag}(m_1,\ldots,m_L)$ is the quark mass matrix
and $A$ is the topological susceptibility in the pure--YM theory.
(This Lagrangian generalizes the one originally proposed in Refs.
\cite{chiral}, which included only the effects due to the anomaly and
the $q\bar{q}$ chiral condensate.)
All the parameters appearing in the Lagrangian must be considered as 
functions of the physical temperature $T$. In particular, the parameters 
$\rho_{\pi}$ and $\rho_X$ determine the expectation values $\langle U \rangle$
and $\langle X \rangle$ and so they are responsible for the behaviour of the
theory respectively across the $SU(L) \otimes SU(L)$ and the $U(1)$ chiral
phase transitions, as follows:
\ba
\rho_\pi|_{T<T_{ch}} &\equiv& {1 \over 2} F_\pi^2 > 0, ~~~
\rho_\pi|_{T>T_{ch}} < 0; \nonumber \\
\rho_X|_{T<T_{U(1)}} &\equiv& {1 \over 2} F_X^2 > 0, ~~~
\rho_X|_{T>T_{U(1)}} < 0.
\label{table}
\ea
The parameter $F_\pi$ is the well--known pion decay constant, while the
parameter $F_X$ is related to the new $U(1)$ axial condensate.
Indeed, from Eq. (\ref{table}),
$\rho_X = {1 \over 2} F_X^2 > 0$ for $T<T_{U(1)}$, and therefore, from Eq.
(\ref{potential}), $\langle X \rangle = F_X/\sqrt{2} \ne 0$. Remembering that
$X \sim {\det} \left( \bar{q}_{sR} q_{tL} \right)$, up to a multiplicative
constant, we find that $F_X$ is proportional to the new $2L$--fermion
condensate $C_{U(1)} = \langle {\cal O}_{U(1)} \rangle$ introduced above.\\
In the same way, the pion decay constant $F_{\pi}$, which controls the breaking
of the $\sgru$ symmetry, is related to the $q\bar{q}$ chiral condensate 
by a simple and well--known proportionality relation (see Refs.
\cite{EM1994a,EM2002} and references therein):
$\langle \bar{q}_i q_i \rangle_{T<T_{ch}} \simeq -{1 \over 2}B_m F_\pi$.
(Moreover, in the simple case of $L$ light quarks with the same mass $m$,
$m^2_{NS} = m B_m/F_\pi$ is the squared mass of the non--singlet pseudoscalar
mesons and one gets the well--known Gell-Mann--Oakes--Renner relation:
$m^2_{NS} F^2_\pi \simeq -2m \langle \bar{q}_i q_i \rangle_{T<T_{ch}}$.)\\
It is not possible to find, in a simple way, the analogous relation between
$F_X$ and the new condensate $C_{U(1)} = \langle {\cal O}_{U(1)} \rangle$.

However, as we have shown in a previous paper \cite{EM2003}, information on
the quantity $F_X$ (i.e., on the new $U(1)$ chiral condensate, to which it is
related) can be derived, in the realistic case of $L=3$ light quarks with
non--zero masses $m_u$, $m_d$ and $m_s$, from the study of the radiative
decays of the pseudoscalar mesons $\eta$ and $\eta'$ in two photons.
In Ref. \cite{EM2003} only the zero--temperature case ($T=0$) has been
considered and a first comparison of our results with the experimental data
has been performed: the results are encouraging, pointing towards a certain
evidence of a non--zero $U(1)$ axial condensate.

In this paper, generalizing the results obtained in Ref. \cite{EM2003},
we study the effects of the $U(1)$ chiral condensate on the radiative decay
$\eta' \to \gamma\gamma$ at finite temperature ($T \ne 0$), so opening the
possibility of a comparison with future heavy--ion experiments.
In Section 2 we first re--discuss the radiative decays of the
pseudoscalar mesons at $T=0$, considering a {\it more general} electromagnetic
anomaly interaction term, obtained by adding a {\it new} electromagnetic
interaction term to the original electromagnetic anomaly term adopted
in Ref. \cite{EM2003} [see Eqs. (\ref{li})--(\ref{li-bar}) below].
As we shall see, the inclusion of this {\it new} electromagnetic interaction
term does not modify, for $T=0$ (or, more generally, for $T<T_{ch}$)
the decay amplitudes for the processes $\pi^0 \to \gamma\gamma$,
$\eta \to \gamma\gamma$ and $\eta' \to \gamma\gamma$:
therefore, all the results (both analytical and numerical) obtained in Ref.
\cite{EM2003}, concerning these processes, remains unaffected.
However, the {\it new} electromagnetic interaction term will prove to be
crucial in the discussion of the $\eta'\to\gamma\gamma$ radiative decay
at finite temperature (in particular for $T>T_{ch}$), which will be studied
in detail in Section 3.

\newsection{Radiative decays of the pseudoscalar mesons at $T=0$}

\noindent
In order to study the radiative decays of the pseudoscalar mesons in two
photons, we have to introduce the electromagnetic interaction in our effective
model (\ref{lagrangian}).
Under {\it local} $U(1)$ electromagnetic transformations:
\be
q \to q' = e^{i\theta e \qu}q, ~~~
A_\mu \to A'_\mu = A_\mu - \partial_\mu \theta,
\ee
the fields $U$ and $X$ transform as follows:
\be
U \to U' = e^{i\theta e \qu} U e^{-i\theta e \qu}, ~~~
X \to X' = X.
\ee
Therefore, we have to replace the derivative of the fields $\partial_\mu U$
and $\partial_\mu X$ with the corresponding {\it covariant} derivatives:
\be
D_{\mu}U=\partial_{\mu}U+ie A_{\mu}[\qu,U], ~~~ D_{\mu}X = \partial_\mu X.
\ee
Here $\qu$ is the quark charge matrix (in units of $e$, the absolute value of
the electron charge):
\be
\label{caricael}
\qu=
\left( \begin{array}{ccc}
\frac{2}{3} & \\
& -\frac{1}{3} \\
& & -\frac{1}{3} \\
\end{array}\right).
\ee
In addition, we have to reproduce the effects of the electromagnetic anomaly,
whose contribution to the four--divergence of the $U(1)$ axial current
$J_{5,\mu}=\bar{q}\gamma_{\mu}\gamma_{5}q$ and of the $SU(3)$ axial currents
$A^{a}_{\mu}=\bar{q}\gamma_{\mu}\gamma_{5}\frac{\tau_{a}}{\sqrt{2}}q$
(the matrices $\tau_a$, with $a=1,\ldots,8$, are the generators of the
algebra of $SU(3)$ in the fundamental representation, with normalization:
$\Tr(\tau_a \tau_b) = \delta_{ab}$) is given by:
\be
(\partial^{\mu}J_{5,\mu})^{e.m.}_{anomaly} = 2\Tr(\qu^{2}) G, ~~~
(\partial^{\mu}A^{a}_{\mu})^{e.m.}_{anomaly} =
2\Tr\left( \qu^{2}\frac{\tau_{a}}{\sqrt{2}}\right) G,
\ee
where $G\equiv\frac{e^{2}N_{c}}{32\pi^{2}}\eps F_{\mu\nu}F_{\rho\sigma}$ 
($F_{\mu\nu}$ being the electromagnetic field--strength tensor), thus 
breaking the corresponding chiral symmetries. We observe that
$\Tr(\qu^{2}\tau_{a}) \ne 0$ only for $a=3$ or $a=8$.\\
We must look for an interaction term ${\cal L}_I$ (constructed with the chiral
Lagrangian fields and the electromagnetic operator $G$) which, under a $U(1)$
axial transformation $q \to q' = e^{-i\alpha\gamma_5}q$, transforms as:
\be
U(1)_A:~~{\cal L}_I \to {\cal L}_I + 2\alpha \Tr(\qu^2)G,
\label{prop-u1}
\ee
while, under $SU(3)$ axial transformations of the type $q \to q' = e^{-i\beta
\gamma_5 \tau_a/\sqrt{2}}q$ (with $a = 3,8$), transforms as:
\be
SU(3)_A:~~{\cal L}_I \to {\cal L}_I + 2\beta \Tr\left( \qu^2
{\tau_a \over \sqrt{2}} \right) G.
\label{prop-su3}
\ee
By virtue of the transformation properties of the fields $U$ and $X$ under a
$\grup$ chiral transformation ($q_L \to V_L q_L$, $q_R \to V_R q_R$
$\Rightarrow$ $U \to V_L U V_R^\dagger$ and $X \to \det(V_L) \det(V_R)^* X$,
where $V_L$ and $V_R$ are arbitrary $3 \times 3$ unitary matrices
\cite{EM1994a,EM2002}), one can see that the most simple term describing the
electromagnetic anomaly interaction term is the following one:
\be
\label{li}
\La_{I}= {i \over 2}G\Tr[\qu^{2}(\ln U-\ln\Ucr)],
\ee
which is exactly the one originally proposed in Ref.
\cite{DiVecchia-Veneziano-et-al.81} and also adopted in Ref. \cite{EM2003}.
However, the presence of the new meson field $X$ allows us to construct
also another electromagnetic interaction term, still proportional to the
pseudoscalar operator $G$, but totally {\it invariant} under $\grup$ chiral
transformations:
\be
\label{delta-li}
\Delta\La_{I} = f_\Delta {i \over 6} G \Tr(\qu^{2})
\left[ \ln (X\det U^\dagger) - \ln (X^\dagger \det U) \right],
\ee
where $f_\Delta$ is an (up--to--now) arbitrary real parameter (the coefficient
$1/6$ has been introduced for convenience: see Section 3).
We can thus add the two expressions (\ref{li}) and (\ref{delta-li})
to form a new (more general) electromagnetic anomaly interaction term
$\overline{\La}_I$, which, of course, satisfies both the transformation
properties (\ref{prop-u1}) and (\ref{prop-su3}), exactly as $\La_{I}$:
\ba
\label{li-bar}
\lefteqn{
\overline{\La}_{I} = \La_{I} + \Delta\La_{I}
= {i \over 2}G\Tr[\qu^{2}(\ln U-\ln\Ucr)] } \nonumber \\
& & + f_\Delta {i \over 6} G \Tr(\qu^{2}) \left[ \ln (X\det U^\dagger)
- \ln (X^\dagger \det U) \right].
\ea
Therefore, we shall consider the following effective chiral Lagrangian, 
which includes the new electromagnetic interaction terms described above:
\ba
\label{lem}
\lefteqn{\La(U,\Ucr,X,\Xcr,Q,A^{\mu})=
\frac{1}{2}\Tr(D_{\mu}UD^{\mu}\Ucr)+
\frac{1}{2}\partial_{\mu}X\partial^{\mu}\Xcr }&& \nonumber\\
& & -V(U,\Ucr,X,\Xcr)+\frac{i}{2}\omega_{1}Q\Tr(\ln U-\ln \Ucr) \nonumber\\
& & +\frac{i}{2}(1-\omega_{1})Q(\ln X-\ln \Xcr)+\frac{1}{2A} Q^{2}
\nonumber\\
& & + \overline{\La}_I -\frac{1}{4}F_{\mu\nu}F^{\mu\nu},
\ea
where the potential term $V(U,\Ucr,X,\Xcr)$ is the one written in Eq. 
(\ref{potential}).\\
The decay amplitude of the generic process ``$meson\to\gamma\gamma$'' is
entirely due to the electromagnetic anomaly interaction term
$\overline{\La}_I$, which can be written more explicitly in terms of the
meson fields $\pi_a$ ($a = 1, \ldots, 8$), $S_\pi$ and $S_X$,
defined as follows \cite{EM1994a,EM1994c,EM2002}:
\ba
U &=& \frac{F_\pi}{\sqrt2}\exp\left[ {i\sqrt{2}\over F_\pi}
\left( \displaystyle\sum_{a=1}^{8}
\pi_{a}\tau_{a}+\frac{S_{\pi}}{\sqrt 3}\I \right) \right],
\nonumber \\
X &=& \frac{F_X}{\sqrt2}\exp\left({i\sqrt{2}\over F_X} S_X\right).
\label{u,x}
\ea
The $\pi_a$ are the self--hermitian fields describing the octet pseudoscalar
mesons; $S_\pi$ is the usual ``quark--antiquark'' $SU(3)$--singlet meson field
associated with $U$, while $S_X$ is the ``exotic'' $6$--fermion meson field
associated with $X$ \cite{EM1994a,EM1994c,EM2002}.\\
Inserting the expressions (\ref{u,x}) into Eq. (\ref{li-bar}), one finds that:
\be
\label{li1}
\overline{\La}_{I}=
-G\frac{1}{3F_{\pi}} \left[ \pi_{3}+\frac{1}{\sqrt{3}}\pi_{8}
+ \frac{2\sqrt{2}}{\sqrt{3}}S_{\pi} -f_\Delta {2\sqrt{2} \over 3 F_X}
(\sqrt{3} F_X S_\pi - F_\pi S_X) \right].
\ee
The fields  $\pi_{3},\pi_{8},S_{\pi},S_{X}$ mix together, while the remaining
$\pi_a$ are already diagonal \cite{EM1994c}.
However, neglecting the experimentally small mass difference between the 
quarks \emph{up} and \emph{down} (i.e., neglecting the experimentally small
violations of the $SU(2)$ isotopic spin), also $\pi_3$ becomes diagonal and can
be identified with the physical state $\pi^0$.
The fields $(\pi_8,S_\pi,S_X)$ can be written in terms of the eigenstates
$(\eta,\eta',\eta_{X})$ as follows:
\be
\pmatrix{ \pi_8 \cr S_\pi \cr S_X } = \mathbf{C}
\pmatrix{ \eta \cr \eta' \cr \eta_X },
\label{diag}
\ee
where $\mathbf{C}$ is the following $3\times3$ orthogonal matrix \cite{EM2003}:
\ba
\label{cambio}
\mathbf{C}=
\left( \begin{array}{ccc}
\alpha_{1} & \alpha_{2} & \alpha_{3}\\
\beta_{1} & \beta_{2} & \beta_{3}\\
\gamma_{1} & \gamma_{2} & \gamma_{3}
\end{array} \right)=
\left(\begin{array}{ccc}
\cos\tilde{\varphi} & -\sin\tilde{\varphi} & 0\\
&&\\
\sin\tilde{\varphi}\,\frac{F_{\pi}}{F_{\eta'}} &
\cos\tilde{\varphi}\,\frac{F_{\pi}}{F_{\eta'}} &
\frac{\sqrt{3}F_{X}}{F_{\eta'}}\\
&&\\
\sin\tilde{\varphi}\,\frac{\sqrt{3}F_{X}}{F_{\eta'}} & 
\cos\tilde{\varphi}\,\frac{\sqrt{3}F_{X}}{F_{\eta'}} & 
-\frac{F_{\pi}}{F_{\eta'}}
\end{array} \right) .
\ea
Here $F_{\eta'}$ is defined as follows \cite{EM2003}:
\be
F_{\eta'} \equiv \sqrt{\csti},
\label{F-eta'}
\ee
and can be identified with the $\eta'$ decay constant in the chiral limit of
zero quark masses. Moreover, $\tilde{\varphi}$ is a mixing angle, which can be
related to the masses of the quarks $m_u$, $m_d$, $m_s$, and therefore to the
masses of the octet mesons, by the following relation \cite{EM2003}:
\be
\label{phitilde}
\tan\tilde{\varphi}
= \frac{F_{\pi}{F_{\eta'}}}{6\sqrt{2}A}(m_{\eta}^{2}-m_{\pi}^{2}),
\ee
where:
$m^{2}_{\pi}=2B\tilde{m}$ and $m_{\eta}^{2}=\frac{2}{3}B(\tilde{m}+2m_s)$,
with: $B \equiv \frac{B_{m}}{2F_{\pi}}$ and
$\tilde{m} \equiv \frac{m_u+m_d}{2}$.\\
Concerning the masses of the two singlet states, we remind that
\cite{EM1994a,EM1994b,EM1994c,EM2002}
the field $\eta'$ has a ``light'' mass, in the sense of the $N_c \to \infty$
limit, being, in the chiral limit of zero quark masses:\footnote{The expression
for the $\eta'$ mass, when including the light--quark masses, reads as follows
\cite{EM1994c}:
$\left( 1 + 3{F_X^2 \over F_\pi^2} \right) m_\eta'^2 + m_\eta^2 -2 m_K^2
= {6A \over F_\pi^2}$,
with: $m_K^2 = B(\tilde{m} + m_s)$.}
\be
m^2_{\eta'} = {6A \over F_\eta'^2} = {6A \over F_\pi^2 + 3F_X^2}
= {\cal O}({1 \over N_c}).
\label{meta'}
\ee
(If we put $F_X=0$, Eq. (\ref{meta'}), or the corresponding expression
including the light--quark masses \cite{EM1994c} reported in the footnote,
reduces to the well--known Witten--Veneziano relation for the $\eta'$ mass
\cite{WV79}.)
On the contrary, the field $\eta_X$ has a sort of ``heavy hadronic'' mass of
order ${\cal O}(N_c^0)$ in the large--$N_c$ limit.
Both the $\eta'$ and the $\eta_X$ have the same quantum numbers (spin, 
parity and so on), but they have a different quark content: one is mostly
$S_\pi \sim i(\bar{q}_{L}q_{R}-\bar{q}_{R}q_{L})$, while the other is mostly
$S_X \sim i[ {\det}(\bar{q}_{sL}q_{tR}) - {\det}(\bar{q}_{sR}q_{tL}) ]$,
as one can see from Eqs. (\ref{diag})--(\ref{cambio}).

The interaction Lagrangian (\ref{li1}), written in terms of the physical fields
$\pi^0,~\eta,~\eta'$ and $\eta_X$, reads as follows: 
\be
\label{limasse}
\overline{\La}_{I}
\equiv-G\frac{1}{3F_{\pi}}\Big(\pi^{0}+a_{1}\,\eta+a_{2}\,\eta'+
\overline{a}_{3}\,\eta_{X}\Big),
\ee
where $a_{i}=\frac{1}{\sqrt3}(\alpha_{i}+2\sqrt2\beta_{i})$
$({\rm for}~i=1,2,3)$, so that:
\ba
\label{aibis}
a_{1}&=&\sqrt{\frac{1}{3}}\Big(\cos\tilde{\varphi}+
2\sqrt2\sin\tilde{\varphi}\,\frac{F_{\pi}}{F_{\eta'}}\Big),\\
a_{2}&=&\sqrt{\frac{1}{3}}\Big(2\sqrt2\cos\tilde{\varphi}\,
\frac{F_{\pi}}{F_{\eta'}}-\sin\tilde{\varphi}\Big),\\
a_{3}&=&2\sqrt2 \Big(\frac{F_{X}}{F_{\eta'}}\Big),
\ea
and, moreover:
\be
\overline{a}_3 = a_3 + \Delta a_3, ~~~ {\rm with:} ~
\Delta a_3 = -f_\Delta {2\sqrt{2} F_{\eta'} \over 3 F_X}.
\ee
The values of the coefficients $a_1$, $a_2$ and $a_3$ are exactly the same
which were calculated in Ref. \cite{EM2003}: therefore, the inclusion of the
new electromagnetic interaction term (\ref{delta-li}) in the expression for
the electromagnetic anomaly interaction term (\ref{li-bar}) only modifies (for
$T=0$ or, more generally, for $T<T_{ch}$: see the discussion in the next
section) the decay amplitude for the process $\eta_X \to \gamma\gamma$, while
leaving unchanged the other decay amplitudes for the processes
$\pi^0 \to \gamma\gamma$, $\eta \to \gamma\gamma$ and $\eta' \to \gamma\gamma$.
Indeed, from Eqs. (\ref{diag}) and (\ref{cambio}) we derive that:
\be
\eta_X = {1 \over F_\eta'}(\sqrt{3}F_X S_\pi - F_\pi S_X),
\label{etax}
\ee
and thus we immediately see that the term proportional to $f_\Delta$ in
Eq. (\ref{li1}) is simply equal to
\be
\Delta\La_I = -G {1 \over 3 F_\pi}
\left( -f_\Delta {2\sqrt{2} F_{\eta'} \over 3 F_X} \right) \eta_X 
= -G {1 \over 3 F_\pi} \Delta a_3 \eta_X.
\ee
The expressions for the decay amplitudes are:
\ba
\label{api0}
A(\pi^{0}\to\gamma\gamma) & = & \frac{e^{2}N_{c}}{12\pi^{2}F_{\pi}}I,\\
\label{aeta}
A(\eta\to\gamma\gamma)
&=&\frac{e^{2}N_{c}}{12\pi^{2}F_{\pi}}\sqrt{\frac{1}{3}}
\Big(\cos\tilde{\varphi}+
2\sqrt2\sin\tilde{\varphi}\,\frac{F_{\pi}}{F_{\eta'}}\Big)I,\\
\label{aeta'}
A(\eta'\to\gamma\gamma)
&=&\frac{e^{2}N_{c}}{12\pi^{2}F_{\pi}}\sqrt{\frac{1}{3}}
\Big(2\sqrt2\cos\tilde{\varphi}\,\frac{F_{\pi}}{F_{\eta'}}-
\sin\tilde{\varphi}\Big)I,\\
\label{aetax}
A(\eta_{X}\to\gamma\gamma)
&=&\frac{e^{2}N_{c}}{12\pi^{2}F_{\pi}}
2\sqrt2 \Big(\frac{F_{X}}{F_{\eta'}} -f_\Delta {F_{\eta'} \over 3 F_X} \Big)I,
\ea
where $I\equiv\varepsilon_{\mu\nu\rho\sigma}
k_{1}^{\mu}\epsilon_{1}^{\nu\ast}k_{2}^{\rho}\epsilon_{2}^{\sigma\ast}$ 
($k_{1}$, $k_{2}$ being the four--momenta of the two final photons and 
$\epsilon_{1}$, $\epsilon_{2}$ their polarizations).
Consequently, the following decay rates (in the real case
$N_c=3$) are derived:
\ba
\label{gammapi0}
\Gamma(\pi^{0}\to\gamma\gamma) &=&
\frac{\alpha^{2}m_{\pi}^{3}}{64\pi^{3}F_{\pi}^{2}}, \\
\label{gammaeta}
\Gamma(\eta\to\gamma\gamma) &=&
\frac{\alpha^{2}m_{\eta}^{3}}{192\pi^{3}F_{\pi}^{2}}\Big(\cos\tilde{\varphi}+
2\sqrt2\sin\tilde{\varphi}\,\frac{F_{\pi}}{F_{\eta'}}\Big)^{2}, \\
\label{gammaeta'}
\Gamma(\eta'\to\gamma\gamma) &=&
\frac{\alpha^{2}m_{\eta'}^{3}}{192\pi^{3}F_{\pi}^{2}}
\Big(2\sqrt2\cos\tilde{\varphi}\,
\frac{F_{\pi}}{F_{\eta'}}-\sin\tilde{\varphi}\Big)^{2}, \\
\label{gammaetax}
\Gamma(\eta_{X}\to\gamma\gamma) &=&
\frac{\alpha^{2}m_{\eta_{X}}^{3}}{8\pi^{3}F_{\pi}^{2}}
\Big(\frac{F_{X}}{F_{\eta'}} -f_\Delta {F_{\eta'} \over 3 F_X}\Big)^{2},
\ea
where $\alpha=e^{2}/4\pi \simeq 1/137$ is the fine--structure constant.\\
The results (\ref{gammapi0})--(\ref{gammaeta'}) are exactly the same which
were found in Ref. \cite{EM2003}. (If we put $F_X=0$, i.e., if we neglect the
new $U(1)$ chiral condensate, the expressions written above reduce to the
corresponding ones derived in Ref. \cite{DiVecchia-Veneziano-et-al.81} using
an effective Lagrangian which includes only the usual $q\bar{q}$ chiral
condensate.) Therefore, also the numerical results obtained in Ref.
\cite{EM2003}, concerning the processes $\eta \to \gamma\gamma$ and
$\eta' \to \gamma\gamma$, remains unaffected. In particular, using the
experimental values for the various quantities which appear in Eqs.
(\ref{gammaeta}) and (\ref{gammaeta'}), i.e.,
\ba
& & F_\pi = 92.4(4) ~{\rm MeV},
\nonumber \\
& & m_\eta = 547.30(12) ~{\rm MeV},
\nonumber \\
& & m_{\eta'} = 957.78(14) ~{\rm MeV},
\nonumber \\
& & \Gamma(\eta\to\gamma\gamma) = 0.46(4) ~{\rm KeV},
\nonumber \\
& & \Gamma(\eta'\to\gamma\gamma) = 4.26(19) ~{\rm KeV},
\label{data}
\ea
we can extract the following values for the quantity $F_X$ and for the
mixing angle $\tilde\varphi$ \cite{EM2003}:
\be
F_X = 27(9) ~{\rm MeV},~~~ \tilde\varphi = 16(3)^0,
\label{results}
\ee
and these values are perfectly consistent with the relation (\ref{phitilde})
for the mixing angle, if we use for the pure--YM topological susceptibility
the estimate $A=(180\pm5~\rm{MeV})^{4}$, obtained from lattice simulations
\cite{lattice}.

Nevertheless, the {\it new} electromagnetic interaction term will play a
crucial role in the discussion of the $\eta'\to\gamma\gamma$ radiative decay
at finite temperature, in particular for $T>T_{ch}$: this will be studied
in detail in the next section.

\newsection{Radiative decays of the pseudoscalar mesons at $T\ne0$}

\noindent
We want now to address the finite--temperature case ($T \ne 0$). As already
said in the Introduction, this will be done (using a sort of mean--field
approximation) simply by considering all the parameters appearing in the
Lagrangian as functions of the physical temperature $T$. In such a way, the
results obtained in the previous section can be extended to the whole region
of temperatures below the chiral transition ($T < T_{ch}$), provided that
the $T$--dependence is included in all the parameters appearing in Eqs.
(\ref{gammapi0})--(\ref{gammaetax}).

What happens when approaching the chiral transition temperature $T_{ch}$
from below ($T \to T_{ch}-$)?
We know that $F_\pi(T) \to 0$ when $T \to T_{ch}-$.
Let us consider, for simplicity, the chiral limit of zero quark masses.
From Eq. (\ref{meta'}) we see that
$m^2_{\eta'} \to {2A(T_{ch}) \over F_X^2(T_{ch})}$ when $T \to T_{ch}-$
and, from Eqs. (\ref{diag})--(\ref{cambio}), we derive:
\be
\eta' = {1 \over F_\eta'}(F_\pi S_\pi + \sqrt{3}F_X S_X),
\label{eta'}
\ee
so that $\eta' \to S_X$ when $T \to T_{ch}-$. In this same limit, the $\eta'$
decay rate (\ref{gammaeta'}) tends to the value:
\be
\label{geta'ft1}
\Gamma(\eta'\to\gamma\gamma) \mathop{\longrightarrow}_{T \to T_{ch}-}
\frac{\alpha^{2}m_{\eta'}^{3}(T_{ch})}{72\pi^{3}F_X^2(T_{ch})}.
\ee
What happens, instead, in the region of temperatures $T_{ch} < T < T_{U(1)}$,
above the chiral phase transition  (where the $SU(3) \otimes SU(3)$ chiral
symmetry is restored, while the $U(1)$ chiral condensate is still present)?
First of all, we observe that we have continuity in the mass spectrum of the
theory through the chiral phase transition at $T=T_{ch}$.
In fact, if we study the mass spectrum of the theory in the region of
temperatures $T_{ch} < T < T_{U(1)}$ \cite{EM1994a,EM1994c,EM2002},
we find that the singlet meson field $S_X$, associated with the 
field $X$ in the chiral Lagrangian, according to the second Eq. (\ref{u,x})
(instead, the first Eq. (\ref{u,x}) is no more valid in this region
of temperatures), has a squared mass given by (in the chiral limit):
$m^2_{S_X} = {2A \over F_X^2}$.
This is nothing but the {\it would--be} Goldstone particle 
coming from the breaking of the $U(1)$ chiral symmetry, i.e., the $\eta'$,
which, for $T>T_{ch}$, is a sort of ``exotic'' matter field of the form
$S_X \sim i[ {\det}(\bar{q}_{sL}q_{tR}) - {\det}(\bar{q}_{sR}q_{tL}) ]$.
Its existence could be proved perhaps in the near future by
heavy--ion experiments.

And what about the $\eta'$ radiative decay rate in the region of
temperatures $T_{ch} < T < T_{U(1)}$?
Since $\eta' = S_X$ above $T_{ch}$,
the electromagnetic anomaly interaction term describing the process
$\eta' \to \gamma\gamma$ for $T>T_{ch}$ is only the part of $\overline{\La}_I$,
written in Eq. (\ref{li-bar}), which depends on the field $X$:
\be
\Delta\La_{S_X\gamma\gamma} = f_\Delta {i \over 6} G \Tr(\qu^{2})
( \ln X - \ln X^\dagger ) = -f_\Delta {2\sqrt{2} \over 9 F_X} G S_X.
\ee
Form this equation we easily derive the following expression for the
$\eta' \to \gamma\gamma$ decay amplitude above $T_{ch}$:
\be
\label{aeta'ft2}
A(\eta'\to\gamma\gamma) \vert_{T>T_{ch}} =
f_\Delta \frac{e^{2}N_{c}\sqrt{2}}{18\pi^{2}F_X}I,
\ee
and, consequently, the following expression for the $\eta' \to \gamma\gamma$
decay rate (in the real case $N_c = 3$) above $T_{ch}$:
\be
\label{geta'ft2}
\Gamma(\eta'\to\gamma\gamma) \vert_{T>T_{ch}} =
f_\Delta \frac{\alpha^{2}m_{\eta'}^{3}}{72\pi^{3}F_X^2}.
\ee
If we require that $\Gamma(\eta'\to\gamma\gamma)$ is a continuous function of
$T$ across the chiral transition at $T_{ch}$, then from Eqs. (\ref{geta'ft1})
and (\ref{geta'ft2}) we obtain the following condition for $f_\Delta$:
\be
f_\Delta (T_{ch}) = 1.
\ee
This means that:
\be
\label{geta'ftch}
\Gamma(\eta'\to\gamma\gamma) \vert_{T=T_{ch}} =
\frac{\alpha^{2}m_{\eta'}^{3}(T_{ch})}{72\pi^{3}F_X^2(T_{ch})}.
\ee
The decay rates and the masses at finite temperature could be determined in
the near--future heavy--ion experiments and then Eq. (\ref{geta'ftch}) will
provide an estimate for the value of $F_X$ at $T=T_{ch}$.
Viceversa, if we were able to determine the value of $F_X$ in some other
independent way (e.g., by lattice simulations: see Ref. \cite{EM2003}),
then Eq. (\ref{geta'ftch}) would give a theoretical estimate of the
ratio $\Gamma(\eta'\to\gamma\gamma)/m_{\eta'}^3$ at $T=T_{ch}$, which
could be compared with the experimental results.
For example, if we make the (very plausible, indeed!) assumption that
the value of $F_X$ does not change very much going from $T=0$ up to
$T=T_{ch}$ (it will vanish at a temperature $T_{U(1)}$ above $T_{ch}$),
i.e., $F_X(T_{ch}) \simeq F_X(0)$, and if we take for $F_X(0)$ the value
reported in Eq. (\ref{results}), then Eq. (\ref{geta'ftch}) furnishes
the following estimate:
\be
\Gamma(\eta'\to\gamma\gamma)\vert_{T=T_{ch}}/m_{\eta'}^3(T_{ch})
= \frac{\alpha^{2}}{72\pi^{3}F_X^2(T_{ch})}
\simeq (3.3^{+4.1}_{-1.4}) \times 10^{-11} ~{\rm MeV}^{-2}.
\ee
In other words, comparing with the corresponding quantities at $T=0$,
reported in Eq. (\ref{data}), one gets that:
\be
{\Gamma(\eta'\to\gamma\gamma)\vert_{T=T_{ch}}/m_{\eta'}^3(T_{ch}) \over
\Gamma(\eta'\to\gamma\gamma)\vert_{T=0}/m_{\eta'}^3(0)} \simeq 7^{+8}_{-3}.
\label{ratio}
\ee
Thus, even with very large errors, due to our poor knowledge of the value of
$F_X$, there is a quite definite prediction that the ratio
$\Gamma(\eta'\to\gamma\gamma)/m_{\eta'}^3$ should have a sharp
increase approaching the chiral transition temperature $T_{ch}$.
(Of course, a smaller value of $F_X$ would result in a larger value for
the ratio in Eq. (\ref{ratio}), and this case seems indeed to be favoured
from the upper limit $F_X \lesssim 20$ MeV obtained from the {\it generalized}
Witten--Veneziano formula for the $\eta'$ mass \cite{EM1994c}.)
One could also argue that it is physically plausible that the $\eta'$ mass
(of the order of $1$ GeV) remains practically unchanged when going from $T=0$
up to $T_{ch}$ (which, from lattice simulations, is known to be of the order
of $170$ MeV: see, e.g., Ref. \cite{Karsch00}): in that case, Eq. (\ref{ratio})
would give an estimate for the ratio between the $\eta'$ decay rates at
$T=T_{ch}$ and $T=0$. However, we want to stress that our result (\ref{ratio})
is more general and does not rely on any given assumption on the behaviour
of $m_{\eta'}(T)$ with the temperature $T$.

\newsection{Conclusions}

\noindent
There are evidences from some lattice results that
a new $U(1)$--breaking condensate survives across the chiral transition
at $T_{ch}$, staying different from zero up to $T_{U(1)} > T_{ch}$.
This fact has important consequences on the pseudoscalar--meson sector,
which can be studied using an effective Lagrangian model, including also
the new $U(1)$ chiral condensate. This model could perhaps be verified in
the near future by heavy--ion experiments, by analysing the pseudoscalar--meson
spectrum in the singlet sector.

In Ref. \cite{EM2003}
we have also investigated the effects of the new $U(1)$ chiral condensate on
the radiative decays, at $T=0$, of the pseudoscalar mesons $\eta$ and $\eta'$
in two photons. A first comparison of our results with the experimental data
has been performed: the results are encouraging, pointing towards a certain
evidence of a non--zero $U(1)$ axial condensate.
In this paper, generalizing the results obtained in Ref. \cite{EM2003}, we have
studied the effects of the $U(1)$ chiral condensate on the radiative decay
$\eta' \to \gamma\gamma$ at finite temperature ($T \ne 0$). In particular,
we have been able to get a quite definite theoretical prediction [see Eq.
(\ref{ratio})] for the ratio between the $\eta'\to\gamma\gamma$ decay rate
and the third power of the $\eta'$ mass in the proximity of the chiral
transition temperature $T_{ch}$ (which, from lattice simulations, is expected
to be of the order of $170$ MeV): this prediction could in principle be tested
in future heavy--ion experiments.

However, as we have already stressed in the conclusions of Ref. \cite{EM2003},
one should keep in mind that our results have been
derived from a very simplified model, obtained doing a first--order expansion
in $1/N_c$ and in the quark masses. We expect that such a model can furnish
only qualitative or, at most, ``semi--quantitative'' predictions.
When going beyond the leading order in $1/N_c$, it becomes
necessary to take into account questions of renormalization--group
behaviour of the various quantities and operators involved in our
theoretical analysis. This issue has been widely discussed in the literature,
both in relation to the proton--spin crisis problem \cite{spin-crisis},
and also in relation to the study of the $\eta,\eta'$ radiative decays
\cite{gamma-gamma}.
Further studies are therefore necessary in order to continue this analysis
from a more quantitative point of view.
We expect that some progress will be done along this line in the near future.

\vfill\eject

{\renewcommand{\Large}{\normalsize}
}

\vfill\eject


\begin{thebibliography}{99}
\bibitem{Bernard-et-al.97}
C. Bernard {\it et al.}, Nucl. Phys. B (Proc. Suppl.) {\bf 53} (1997) 442;\\
C. Bernard {\it et al.}, Phys. Rev. Lett. {\bf 78} (1997) 598.
\bibitem{Karsch00}
F. Karsch, Nucl. Phys. B (Proc. Suppl.) {\bf 83--84} (2000) 14.
\bibitem{Vranas00}
P.M. Vranas, Nucl. Phys. B (Proc. Suppl.) {\bf 83--84} (2000) 414.
\bibitem{EM1994a}
E. Meggiolaro, Z. Phys. C {\bf 62} (1994) 669.
\bibitem{EM1994b}
E. Meggiolaro, Z. Phys. C {\bf 62} (1994) 679.
\bibitem{EM1994c}
E. Meggiolaro, Z. Phys. C {\bf 64} (1994) 323.
\bibitem{EM2002}
E. Meggiolaro, \emph{``Remarks on the $U(1)$ axial symmetry in QCD at zero and
non--zero temperature''}, preprint IFUP--TH/2002--24; hep--ph/0206236.
\bibitem{tHooft76}
G. 'tHooft, Phys. Rev. Lett. {\bf 37} (1976) 8;\\
G. 'tHooft, Phys. Rev. D {\bf 14} (1976) 3432.
\bibitem{EM1995}
A. Di Giacomo and E. Meggiolaro, Nucl. Phys. B (Proc. Suppl.) {\bf 42}
(1995) 478.
\bibitem{chiral}
P. Di Vecchia and G. Veneziano, Nucl. Phys. B {\bf 171} (1980) 253;\\
E. Witten, Annals of Physics {\bf 128} (1980) 363;\\
C. Rosenzweig, J. Schechter and C.G. Trahern,
Phys. Rev. D {\bf 21} (1980) 3388;\\
P. Nath and R. Arnowitt, Phys. Rev. D {\bf 23} (1981) 473;\\
K. Kawarabayashi and N. Ohta, Nucl. Phys. B {\bf 175} (1980) 477.
\bibitem{EM2003}
M. Marchi and E. Meggiolaro, Nucl. Phys. B {\bf 665} (2003) 425.
\bibitem{DiVecchia-Veneziano-et-al.81}
P. Di Vecchia, F. Nicodemi, R. Pettorino and G. Veneziano, Nucl. Phys. B
{\bf 181} (1981) 318.
\bibitem{WV79}
E. Witten, Nucl. Phys. B {\bf 156} (1979) 269;\\
G. Veneziano, Nucl. Phys. B {\bf 159} (1979) 213.
\bibitem{lattice}
M. Teper, Phys. Lett. B {\bf 202} (1988) 553;\\
M. Campostrini, A. Di Giacomo, Y. G\"unduc, M.P. Lombardo, H. Panagopoulos and
R. Tripiccione, Phys. Lett. B {\bf 252} (1990) 436;\\
B. All\'es, M. D'Elia and A. Di Giacomo, Nucl. Phys. B {\bf 494} (1997) 281.
\bibitem{spin-crisis}
G.M. Shore and G. Veneziano, Phys. Lett. B {\bf 244} (1990) 75;\\
G.M. Shore and G. Veneziano, Nucl. Phys. B {\bf 381} (1992) 23;\\
S. Narison, G.M. Shore and G. Veneziano, Nucl. Phys. B {\bf 433} (1995) 209;\\
S. Narison, G.M. Shore and G. Veneziano, Nucl. Phys. B {\bf 546} (1999) 235.
\bibitem{gamma-gamma}
G.M. Shore and G. Veneziano, Nucl. Phys. B {\bf 381} (1992) 3;\\
G.M. Shore, Nucl. Phys. B (Proc. Suppl.) {\bf 86} (2000) 368;\\
G.M. Shore, Nucl. Phys. B {\bf 569} (2000) 107;\\
G.M. Shore, Phys. Scripta T {\bf 99} (2002) 84.
\end{thebibliography}
\end{document}